\documentclass[referee]{raa}
\usepackage{graphicx,times}
\usepackage{natbib}
\usepackage{amssymb,amsmath}
\bibpunct{(}{)}{;}{a}{}{,}

\usepackage[a4paper=true,dvipdfm=true,pagebackref=true]{hyperref}
\hypersetup{pdftitle = The title of my PDF, pdfauthor = My name, pdfsubject= The subject, pdfkeywords = keyword1 keyword2 keyword3}
\hypersetup{colorlinks = true, linkcolor = green, anchorcolor = red, citecolor = blue, filecolor = red, pagecolor = red, urlcolor = red}

\begin{document}

   \title{Study of grain alignment efficiency and a distance estimate for
small globule CB4}

 \volnopage{ {\bf 2014} Vol.\ {\bf X} No. {\bf XX}, 000--000}
   \setcounter{page}{1}

   \author{Ajoy Barman and Himadri Sekhar Das
   }

   \institute{Department of Physics, Assam University, Silchar 788011, India; {\it hsdas@iucaa.ernet.in}\\
\vs \no
   {\small Received 2014 Nov XX; accepted 2014 XXXX}
}

\abstract{We study the polarization efficiency (defined as the ratio of polarization
to extinction) of stars in the background of the small, nearly spherical and isolated
Bok globule CB4 to understand the grain alignment process. A decrease in polarization
efficiency with an increase in visual extinction is noticed. This suggests that the
observed polarization in lines of sight which intercept a Bok globule tends to show
dominance of dust grains in the outer layers of the globule. This finding is consistent
with the results obtained for other clouds in the past. We determined the distance to
the cloud CB4 using near-infrared photometry (2MASS JHKS colors) of moderately
obscured stars located at the periphery of the cloud. From the extinction-distance plot,
the distance to this cloud is estimated to be ($459 \pm 85$) pc.
\keywords{polarization –- ISM: clouds –- dust, extinction –- ISM: individual objects: CB4.
}
}

   \authorrunning{Barman \& Das }            
   \titlerunning{Study of Grain Alignment Efficiency and a Distance Estimate for Small Globule CB4}  
   \maketitle

%
\section{Introduction}

Linear polarization measurement of background field stars in a star forming cloud has been used to
study the large scale structure of the interstellar magnetic field (\cite{h}; \cite{flp}; \cite{bwd}). The efficiency with which polarization is produced by dust grains is sensitive to physical conditions of the cloud, with a tendency to drop systematically with increasing extinction or optical depth inside interstellar clouds (\cite{gjl2}; \cite{gwl}; \cite{whl}; \cite{cdp}). This observation suggests that the observed polarization in a particular direction or line of sight which intercept a dark cloud show dominance of dust grains in the outer layers of the cloud. However background stars observed through dense region of the cloud show smaller degree of polarization which suggests much lower dust columns in the observed lines of sight. This also suggests that the polarization studies of background stars in a cloud can well explain the external magnetic field, but it can not provide information on magnetic field structure within the cloud (\cite{w1}).

To model the observed variation in polarization with extinction along different lines of sight, \cite{j} and \cite{jkd} studied this in terms of a magnetic field with distinct random and uniform components. It has been observed by different investigators that the polarization efficiency decreases with increase in extinction (\cite{tns}; \cite{gjl1}; \cite{gwl}; \cite{whl}; \cite{cdp}). This can be represented by a power law $p/A_V \propto A_V^{-\alpha}$ for some power law index $\alpha$. From the study of dust grain alignment in Taurus dark cloud, \cite{gwl} found a strong correlation between polarization efficiency and extinction, which was fitted by power law $p/A_V \propto A_V^{-0.56}$, in the magnitude region ($0\leq A_V \leq 25$). \cite{whl} also studied the polarization efficiency toward molecular clouds with an aim to understand grain alignment mechanisms in dense regions of the interstellar medium and found an empirical relation: $p_{\lambda}/\tau_{\lambda} \propto [A_V]^{-0.52}$ (where $\tau_{\lambda}$ is the optical depth). They did not find any significant change in the behavior in the transition region between the diffuse outer layers and dense inner regions of clouds. Recently, \cite{cdp} studied the polarization efficiencies in three selected Bok globules CB56, CB60 and CB69. They found a trend of decreasing polarization efficiency with increasing extinction which can be well represented by a power law, $p_V/A_V \propto A_V^{-\alpha}$, where $\alpha = -0.56\pm0.36, -0.59\pm0.51$ and $-0.52\pm0.49$ for CB56, CB60 and CB69, respectively. To study the alignment of grains, radiative torque mechanism has been used by researchers to calculate the expected relationship between polarization efficiency and extinction through a homogeneous cloud (\cite{dm}; \cite{dw1}; \cite{dw2}; \cite{lh}; \cite{whl}; \cite{hl}). \cite{hl} studied grain alignment by radiative torques for various environment conditions, including interstellar medium, dense molecular clouds and accretion discs.

The determination of distance to an interstellar cloud help researchers to obtain luminosities of the protostars embedded in the clouds (\cite{yc}) and to estimate the sizes, masses and densities of the cloud (\cite{cyh}). Several methods have been adopted to estimate the distance to cloud. These are  Wolf diagrams method (\cite{w2}), star counts method (\cite{bb}), spectroscopic method (\cite{hbm}; \cite{hfa}), photometric method (\cite{s}; \cite{pc}; \cite{mb}; \cite{mmb}; \cite{mlb}) etc. \cite{mb} determined the distances to nine dark globules by a method which uses optical and near-infrared photometry of stars projected towards the field containing the globules. Recently, \cite{emp} determined the distance to starless dark cloud LDN 1570 using near-IR photometry from 2MASS.

In this work, we plan to study the polarization efficiency of field stars background to the isolated Bok globule CB4. We also determined the distance to the cloud CB4 using a technique developed by \cite{mlb} which uses JHK near-infrared photometry.

\section{Object}
\subsection{CB4}
CB4 is a small, nearly spherical and isolated Bok globule situated at a distance of 600 parsec (\cite{dc}).
CB4 is situated far away from the galactic plane ($b \approx - 10^{\circ}$). The angular size of this cloud on the POSS maps is $2' \times 1'$ and the position angle of the major axis (measured east from north) is $45^{\circ}$. An IRAS point source is located at southern edge of the cloud which is detected at $100\mu m$ only (\cite{cb}). No further evidence of an embedded source has been observed up to 160$\mu m$. CB4 has been spectroscopically studied in CO and its isotopes (\cite{dc}, \cite{cyh}). The observations have shown that the globule velocity is about $-$11.3 km s$^{-1}$ with the velocity of dispersion 0.57 km s$^{-1}$. Furthermore, the study of deep IRAS aperture photometry and molecular line showed that CB4 is cold ($\approx$ 7K) and does not contain a dense core ($n > 10^4 $cm$^{-3}$) (\cite{cyh}, \cite{kcm} and \cite{tpm}).

\section{Data}
\subsection{Polarimetric data}
The polarimetric data of 80 field stars of CB4 at V-band has been taken from \cite{kcl}. Four fields were initially observed by them to cover a region of $\sim 8' \times 8'$ surrounding CB4. Two other fields were also observed again with a slightly larger pixel size and using longer integrations. The histogram of polarization position angles (PA) of 80 field stars displayed a strong peak around mean PA 65$^\circ$ with a dispersion of 13$^\circ$. This position angle represents the preferred magnetic field direction and the narrow width of the peak indicates a larger uniform field. The unweighted mean polarization and weighted mean polarization were calculated by Kane et al. to be (2.84$\pm$0.25)\% and (0.96$\pm$0.07)\% which indicate a large uniform magnetic field.

\subsection{Photometric data}
The near IR photometric data ($J$, $H$ and $K_S$) of the field stars of CB4 has been taken from the 2MASS All-Sky Catalog of Point Sources (\cite{cs}) which satisfies the following criteria:
\begin{enumerate}
  \item photometric uncertainty $\sigma_J$, $\sigma_H$, $\sigma_{K_S} \le 0.035$  in all three filters,
  \item signal-to-noise ratio (SNR) $>$ 10 which is shown by photometric quality flag of ``AAA" in all three filters.
\end{enumerate}

The position of 80 stars are shown by RA(2000) and DEC(2000) shown in Table 2 of \cite{kcl}. It is to be noted that we will designate all field  stars by its serial number used in \cite{kcl}'s paper and this numbering has been shown in column number two of Table 1. The photometric magnitude of each field stars has been searched via CDS's VizieR service, using 2 arcmin search radii as query. The $J$, $H$ and $K_S$ magnitudes of 80 stars are available in the catalogue. We also checked the SIMBAD astronomical database to search for any variable star out of 80 field stars, but we could not find any variable star. We found that only 60 stars satisfy the above criteria and our analysis is restricted to these stars only. Further we could not find the uncertainties of magnitudes of six stars in the 2MASS catalogue which are star number 4, 20, 27, 50, 51 and 74. So we discarded them from our analysis. Thus total number of stars considered for our analysis is 54.

\section{Results}

\subsection{Spectral types and $A_V$}
We used the technique developed by \cite{mlb} to infer spectral type, and hence absolute magnitude and intrinsic colors of normal main sequence stars and giants from NIR photometry.  This method gives distances to globules that are relatively close ($\lesssim$ 500 pc) with a precision of $\sim 20\%$. In this technique, the fields containing the cores are divided into small sub-fields to avoid complications created by the wrong classifications of giants into dwarfs. Actually the increase in the extinction caused by a cloud should occur almost at the same distance in all the fields, if the whole cloud were located at the same distance, the wrongly classified stars in the sub-fields would show high extinction not at the same but at random distances (\cite{mlb}). But for small and isolated clouds, it would be difficult to divide the field containing the cloud into sub-fields with a sufficient number of stars projected on to them. Therefore the distance measurement is somehow hard for CB4 as it is a small, nearly spherical and isolated Bok globule. It would be difficult to subdivide the field with only limited number of stars. So we consider a single field surrounding the core of the CB4.

In this technique, stars with $(J-K_S)\le 0.75$ were selected from the fields containing the cloud  to eliminate M-type stars from the analysis. Then a set of dereddened colors $(J-H)_o$ and $(H-K_S)_o$ has been produced for each star from their observed colors $(J-H)$ and $(H-K_S)$ using trial values of $A_V$ and a normal interstellar extinction law (i.e., total-to-selective extinction value, $R_V = 3.1$) in the equations (\cite{rl}):
\begin{eqnarray}
    (J-H)_o &=& (J-H) - 0.107\times A_V\\
    (H-K_S)_o &=& (H-K_S) - 0.063\times A_V
\end{eqnarray}

The maximum extinction that could be computed using this method is limited to $A_V \approx 4$ mag. The trial values of $A_V$ is chosen in the range $0-10$ mag with a step size of 0.01 mag. The calculated set of dereddened color indices are then compared with the intrinsic color indices $(J-H)_{in}$ and $(H-K_S)_{in}$ of normal main sequence stars and giants, produced using the procedures discussed in \cite{mlb}. The best fit values of the dereddened colors to the intrinsic colors giving a minimum value of $\chi^2$ then give the corresponding spectral type and $A_V$ for the star. The solutions which give $\chi^2_{min} \le 0.1$ is considered for this analysis.

The uncertainty in $A_V$ is given by
\begin{equation}
\sigma(A_V) = \sqrt{4.7^2.\sigma^2_{JH} + 7.9^2.\sigma^2_{HK_S} + 2\times37\times cov(JH,HK_S)}
\end{equation}
where $\sigma^2_{JH} = \sigma^2_{J}+\sigma^2_{H}$, $\sigma^2_{HK_S} = \sigma^2_{H}+\sigma^2_{K_S}$ and $cov(JH, HK_S) = r_s \times \sigma_{JH} \sigma_{HK_S}$. The Spearman rank-order correlation coefficient ($r_s$) is calculated from uncertainties in $(J-H)$ and $(H-K_S)$ colors which shows a strong correlation between them.  The maximum uncertainty in the $A_V$ in our case is estimated to be $0.5$ mag.

 We found that 14 stars have $(J-K_S)>0.75$, so we also discarded them from our analysis. We used this technique to estimate the $A_V$ and absolute magnitudes for 40 stars. It has been found that only 20 stars have $\chi^2_{min} \le 0.1$ for which it is possible to infer the spectral type and the $A_V$.  The results are shown in Table-1.

 In \cite{mlb}'s technique, the field stars in the globule are assumed to be normal main sequence stars and therefore the luminosity class V has been assigned to all stars. However the presence of non main sequence stars may result wrong distance to stars. The intrinsic color and luminosity may be reliable if the spectral type is correctly determined and happens to be main sequence. For this we need to plot $(B-V)$ vs $(V-R)$ color - color diagram. The photometric data ($BVR$) of the field stars of CB4 has been taken from the VizieR database of astronomical catalogues, namely, UCAC4 (\cite{zfg}) and NOMAD (\cite{zml}). The $B, V, R$  data of star with serial number 79 is not available in the catalogue. In Fig. 1, color - color diagram along with error bars for 19 field stars of CB4 has been plotted. It can be seen from the graph that most of the field stars are
clustered together along a diagonal locus. These diagonally clustered
stars in the colour - colour diagram represent the main-sequence
stars, while the scattered ones (Star \# 32, 66 and 76) represent non-main-sequence stars.
So we discarded three stars from our analysis. To calculate $r_s$, we plotted uncertainties in $(J-H)$ and $(H-K_S)$ colors for 16 stars which is shown in Fig.2. The value is found to be 0.95. The results obtained from this technique are shown in Table-2.

\begin{figure*}[h]
\begin{center}
\includegraphics[width=90mm]{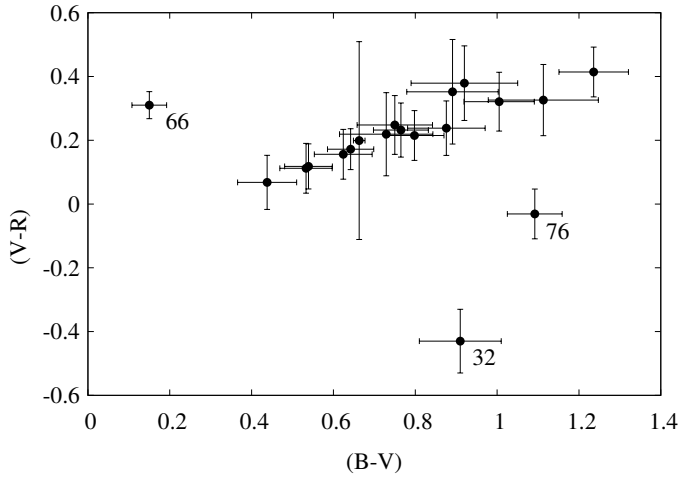}
 \vspace{0.5cm}
\caption{Plot of $B - V$ versus $V - R$ colour - colour diagram of 19 field stars
in the CB4. Diagonally clustered points represent
stars that are in the main sequence. The stars that do not belong to the main
sequence are represented by their serial numbers as mentioned in Table 1.}
\end{center}
\end{figure*}

\begin{figure*}[h]
\begin{center}
\includegraphics[width=90mm]{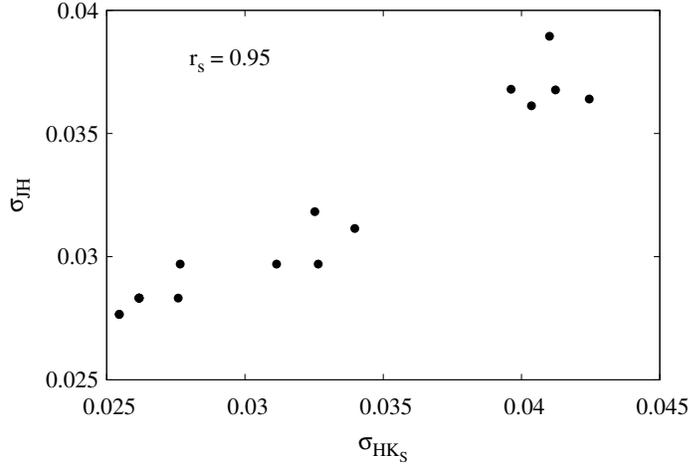}
 \vspace{0.5cm}
\caption{The relation between the $\sigma_{HK_S}$ and $\sigma_{JH}$ of the 16 stars. The $r_s$ value indicates a strong correlation between the uncertainties of the two colors.}
\end{center}
\end{figure*}

{
\renewcommand{\baselinestretch}{1.5}
\small\normalsize
\begin{table}
\begin{center}
\caption{The $J$, $H$ and $K_S$ magnitudes of 20 stars along with  spectral type, absolute magnitude ($M_{K_S}$) and extinction ($A_v$). The second column refers to the star number which corresponds to the serial number in table 2 of \cite{kcl}. }
{\small
\begin{tabular}{c c c c c c c c c}
\hline
S/N & Star   & $J$  & $H$  & $K_S$ & Sp Type   & $M_{K_S}$ & $A_V$\\
    &        & (mag)& (mag)&(mag)  &           &           &      \\
\hline
1	&	1	&		10.942$\pm$0.021	&	10.604$\pm$0.018	&	10.533$\pm$0.018	&	G4V	&	3.45	&	0  \\ \hline																					 
2	&	9	&		11.026$\pm$0.021	&	10.461$\pm$0.019	&	10.305$\pm$0.018	&	K4V	&	4.22	&	0.40 \\ \hline																					 
3	&	11	&		11.888$\pm$0.021	&	11.412$\pm$0.019	&	11.282$\pm$0.018	&	K2V	&	4.15	&	0.23\\ \hline																					 
4	&	14	&		11.555$\pm$0.021	&	11.185$\pm$0.018	&	11.099$\pm$0.018	&	G9V	&	4.03	&	0 \\ \hline																					 
5	&	17	&		11.653$\pm$0.021	&	11.186$\pm$0.019	&	11.1$\pm$0.018	&	K1V	&	4.13	&	0  	\\ \hline																					 
6	&	18	&		12.715$\pm$0.022	&	12.515$\pm$0.023	&	12.396$\pm$0.023	&	A8V	&	1.64	&	1.05 \\ \hline																					 
7	&	23	&		13.083$\pm$0.021	&	12.702$\pm$0.023	&	12.607$\pm$0.025	&	K0V	&	4.15	&	1.34  	\\ \hline																					 
8	&	26	&		11.52$\pm$0.021	&	11.206$\pm$0.019	&	11.108$\pm$0.018	&	G3V	&	3.34	&	0.24  \\ \hline																					 
9	&	32	&		12.978$\pm$0.022	&	12.552$\pm$0.023	&	12.382$\pm$0.023	&	G2V	&	3.29	&	1.44 \\ \hline																					 
10	&	35	&		10.335$\pm$0.021	&	10.199$\pm$0.021	&	10.127$\pm$0.018	&	F0V	&	1.86	&	0.16  	\\ \hline																					 
11	&	36	&		13.332$\pm$0.022	&	13.036$\pm$0.029	&	12.98$\pm$0.031	&	G2V	&	3.29	&	0	 	\\ \hline																					 
12	&	41	&		12.897$\pm$0.026	&	12.354$\pm$0.029	&	12.268$\pm$0.029	&	K3V	&	4.15	&	0 \\ \hline																					 
13	&	52	&		11.055$\pm$0.021	&	10.863$\pm$0.019	&	10.773$\pm$0.02	&	F1V	&	2.00	&	0.45 	 \\ \hline																					 
14	&	61	&		12.691$\pm$0.021	&	12.418$\pm$0.021	&	12.321$\pm$0.025	&	F7V	&	2.75	&	0.44 \\ \hline																					 
15	&	62	&		13.038$\pm$0.025	&	12.693$\pm$0.027	&	12.566$\pm$0.029	&	G0V	&	3.19	&	0.81 	\\ \hline																					 
16	&	66	&		13.19$\pm$0.025	&	12.865$\pm$0.031	&	12.789$\pm$0.034	&	G4V	&	3.45	&	0	\\ \hline																					 
17	&	71	&		12.091$\pm$0.021	&	11.881$\pm$0.021	&	11.783$\pm$0.023	&	F1V	&	2.00	&	0.59 	\\ \hline																					 
18	&	76	&		12.551$\pm$0.022	&	12.238$\pm$0.023	&	12.104$\pm$0.024	&	F6V	&	2.62	&	0.98  	\\ \hline																					 
19	&	77	&		13.3$\pm$0.024	&	12.9$\pm$0.027	&	12.784$\pm$0.03	&	G9V	&	4.03	&	0.35 	\\ \hline																					 
20	&	79	&		13.483$\pm$0.026	&	13.103$\pm$0.026	&	13.037$\pm$0.032	&	G9V	&	4.03	&	0  	\\ \hline
\end{tabular}
}
\end{center}
\end{table}
}

{
\renewcommand{\baselinestretch}{1.5}
\small\normalsize
\begin{table}
\begin{center}
\caption{Polarization ($p$), extinction ($A_V$), polarization efficiency ($p/A_V$) and distance (in parsec) of 16 background stars in CB4. The second column refers to the star number which corresponds to the serial number in table 2 of \cite{kcl}. }
{\small
\begin{tabular}{c c c c c c }
\hline
S/N & Star & $p$ & $A_V$ & $p/A_v$           & $d$ \\
    &      &  (\%) &     & (\% mag$^{-1})$   &  (pc)\\
\hline
1	&	1	& 0.30$\pm$0.05	&	0 & -	& 261$\pm$48 \\ \hline																					
2	&	9	& 1.66$\pm$0.41	&	0.40 & 4.15$\pm$0.53 & 161$\pm$30 \\ \hline																					
3	&	11	&1.26$\pm$0.38	&	0.23	& 5.48$\pm$0.51 & 264$\pm$49 	\\ \hline
4	&	14	&0.49$\pm$0.08	&	0 & -	& 259$\pm$48 	\\ \hline																					
5	&	17	&2.85$\pm$0.38	&	0 & -	& 248$\pm$46 	\\ \hline																					
6	&	18	&1.52$\pm$0.06	&	1.05 & 1.45$\pm$0.41	& 1342$\pm$249 	\\ \hline													
7	&	23	&1.37$\pm$0.12	&	1.34 & 1.02$\pm$0.43	& 459$\pm$85 	\\ \hline																				
8	&	26	&1.07$\pm$0.04	&   0.24 & 4.46$\pm$0.34	& 353$\pm$65 	 \\ \hline																				
9	&	35	&0.27$\pm$0.07	&	0.16 & 1.69$\pm$0.36	& 446$\pm$83 	\\ \hline																				
10	&	36	&1.16$\pm$0.15	&	0	& - & 867$\pm$162 	\\ \hline																					
11	&	41	&1.94$\pm$0.23	&	0 & -	& 420$\pm$78 	\\ \hline																					
12	&	52	&0.97$\pm$0.12	&	0.45& 2.16$\pm$0.37	& 555$\pm$103 	 \\ \hline																					
13	&	61	&0.47$\pm$0.08	&	0.44 & 1.07$\pm$0.40	& 802$\pm$149 	\\ \hline																				
14	&	62	&1.14$\pm$0.27	&	0.81 & 1.41$\pm$0.55	& 720$\pm$134 	\\ \hline																				
15	&	71	&0.55$\pm$0.12	&	0.59 & 0.93$\pm$0.40	& 878$\pm$163 	\\ \hline																				
16	&	77	&1.51$\pm$0.25	&	0.35& 4.31$\pm$0.54	& 553$\pm$103 	\\ \hline																					
\end{tabular}
}
\end{center}
\end{table}
}

\subsection{Polarization efficiency}
The study of the variation of background star polarization with extinction ($A_V$) and polarization efficiency (defined as the ratio of $p$ to $A_V$) can provide useful information regarding the nature of dust and the magnetic field associated with the cloud (\cite{gwl}; \cite{agbms}). The polarization efficiency depends on both the properties of the dust grains and the degree of alignment of the grains. \cite{w} calculated the theoretical upper limit which is given by $p/A_v \le 14\%$ mag$^{-1}$. However, the observational upper limit of $p/A_v$ is found to be $\sim 3 \%$ mag$^{-1}$, a factor of four less than the predicted value for the ideal scenario.

In this work, we will study the polarization efficiency of background field stars of CB4 to investigate the alignment of non spherical dust grains. It is to be noted that we designated all field  stars by its serial number used in \cite{kcl}'s paper and this numbering has been shown in column number two of Table 1. The extinction value and polarization efficiency of background stars are shown in Table 2. In Fig. 3a, we plotted $A_V$ versus $p$ for all the field stars in the cloud CB4. The solid line in the figure represents optimum alignment in the general interstellar medium according to \cite{smf}: $p_{\textrm{max}}/A_V = 3$ percent mag$^{-1}$, where $R_V = 3.1$. It can be seen that only four stars located outside the limit and majority of stars lie below the solid line which suggests that the characteristics of material composing CB4 is consistent with those of material found in the diffuse interstellar medium (ISM).

In Fig. 3b, we plotted $A_V$ vs $p/A_V$ for cloud CB4. The plot shows that the polarization drops with increasing extinction.  Thus background stars observed through dense region of the cloud show smaller degree of polarization suggesting much lower dust columns in the observed lines of sight. This suggests that the observed polarization  at optical region in a particular line of sight which intercept a dark cloud show dominance of dust grains in the outer layers of the cloud.  Recently, \cite{cdp} found a trend of decreasing polarization efficiency with increasing extinction which can be well represented by a power law, $p_V/A_V \propto A_V^{-\alpha}$, where $\alpha = -0.56\pm0.36, -0.59\pm0.51$ and $-0.52\pm0.49$ for Bok globules CB56, CB60 and CB69, respectively. It is to be noted that the study of  polarization efficiency at optical region can give us information about the grain alignment in outer layers of the cloud.  The alignment of grains by radiative torque has been used by researchers to calculate the expected relationship between polarization efficiency and extinction through a homogeneous cloud (\cite{dm}; \cite{dw1}; \cite{dw2}; \cite{lh}; \cite{whl}; \cite{hl}).

\begin{figure}
\vspace{5cm}
  \centering
  \includegraphics[scale=1, angle=0]{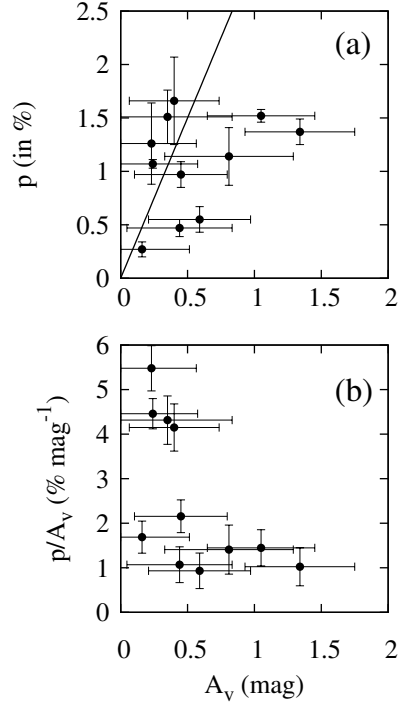}
  \vspace{1cm}
   \caption{(a) Plot of $A_V$ versus $p$ for 11 field stars. The solid line in the figure represents optimum alignment in the general interstellar medium according to \cite{smf}: $p_{\textrm{max}}/A_V = 3$ percent mag$^{-1}$, where $R_V = 3.1$. It can be seen that four stars located outside the limit. (b) plot of $A_V$  versus $p/A_V $  (where $R_V = 3.1$) for 11 background stars (see Table 2).}\label{figexample}
    \end{figure}

\subsection{Estimation of distance}
The photometric distance $d$ to a star is estimated using equation:
\begin{equation}
d\; \textrm{(in pc)} = 10^{(K_S-M_{K_S}+5-A_{K_S})/5}
\end{equation}
where $K_S$, $M_{K_S}$ and $A_{K_S}$ are apparent magnitude, absolute magnitude and extinction, respectively. The distance to the cloud is typically estimated from the first star that shows a sharp peak in extinction in an $A_V$ vs. $d$ plot. It is to be noted that $A_{K_S} = 0.112\times A_V$.

The uncertainty in distance is estimated using the expression (\cite{mlb}),
\begin{equation}
    \sigma_d = \sqrt{(\sigma^2_{K_S} + \sigma^2_{M_{K_S}} + \sigma^2_{A_{K_S}})\times (d/2.17)^2}
\end{equation}
where $\sigma_{K_S}$ is the uncertainty in $K_S$ band, $\sigma_{M_{K_S}}$ is the uncertainty in the estimation of the absolute magnitude and $\sigma_{A_{K_S}}$ is the uncertainty in the $A_{K_S}$ estimated by the method. The uncertainty in the $\sigma_{M_{K_S}}$ is assumed to be 0.4 while calculating $\sigma_d$ in the distances for all the stars.  The typical errors in our distance estimates for the field stars in CB4 are $\sim 19\%$.


\begin{figure}
  \centering
  \includegraphics[width=90mm]{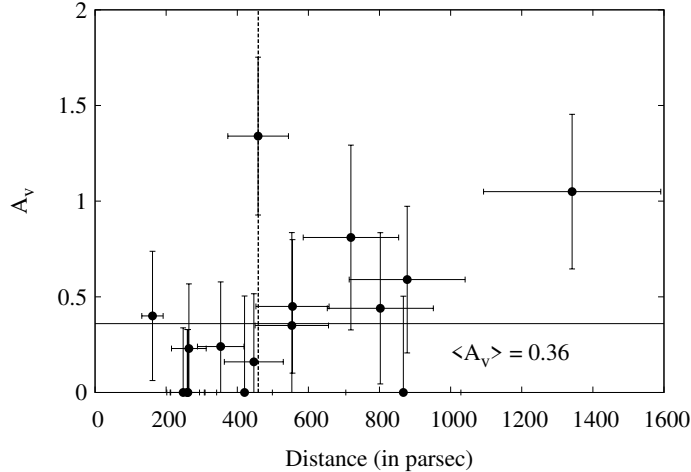}
   \caption{Extinction ($A_V$) versus distance for stars in the vicinity of the CB4. The average extinction ($<A_V>$) of all the field stars is  0.36 and is shown by solid line. The vertical dashed line is drawn at a distance of 459 parsec where sudden rise in the $A_V$ occurs.}\label{figexample}
    \end{figure}

The extinction ($A_V$) versus distance ($d$) plot has been extensively used by several investigators (\cite{dc}; \cite{kdh}; \cite{pc}; \cite{mmb}; \cite{mlb} etc.) to determine the distance of globules. It is noticed that the extinction of stars located in front of the dark globule shows almost zero reddening value. The presence of cloud shows sudden rise of extinction at a certain distance. The distance to the cloud is typically estimated from the first star that shows a sharp peak in $A_V$. We notice from  Fig. 4  that star number  23 shows a sudden rise of extinction and then a fall in extinction is observed. Thus the distance of the cloud is  ($459 \pm 85$) parsec. Using the foreground star method and color excess approaches, \cite{dc} estimated distance to the CB4 to be $\sim 500-700$pc. They therefore adopted 600 pc as the distance to this globule. The distance determination using JHK near-infrared photometry is  slightly smaller than the distance derived by \cite{dc}.  However, the present analysis is
restricted to a limited number of data points.

\section{Summary}

\begin{enumerate}
  \item The polarization efficiency of CB4 is studied and shows a decrease in polarization efficiency
with an increase in extinction along the observed line of sight. This finding is consistent with
the results obtained for other clouds in the past.

  \item The method for determining distances to Bok globules using the NIR photometric technique has
been described. The plot of $A_V$ versus $d$ has been made and the distance at which $A_V$ shows a
sudden rise of extinction is taken as the distance to the cloud. From the study of the distribution
of reddening, we estimate a value of ($459 \pm 85$) pc for the distance to CB4.
\end{enumerate}


\section{Acknowledgement}
The authors sincerely acknowledge the anonymous referee for constructive comments which definitely helped to improve the quality of the paper. This work makes use of data products from the Two Micron All Sky Survey (2MASS),
which is a joint project of the University of Massachusetts and
the Infrared Processing and Analysis Center/California Institute
of Technology, funded by the National Aeronautics and Space
Administration and the National Science Foundation. We also acknowledge the use of the VizieR
data base of astronomical catalogues, namely UCAC4 \citep{zfg} and NOMAD \citep{zml}. This research has also made use of the SIMBAD
database, operated at CDS, Strasbourg, France. This work is supported by the Science and Engineering Research Board (SERB), a statutory body under Department of Science and Technology (DST), Government of India, under Fast Track scheme for Young Scientist (SR/FTP/PS-092/2011).

\end{document}